\documentclass[11pt]{article}
\usepackage{amssymb}
\usepackage{subfig}
\usepackage{color}
\usepackage{epsfig,amssymb,amsfonts,amsmath,graphicx,dsfont,cite,xfrac}
\usepackage{authblk}
\usepackage{ulem}
\definecolor{mygray}{gray}{0.5}
\usepackage{cite}
\usepackage[colorlinks=true,linkcolor=blue,citecolor=red]{hyperref}
\usepackage{multirow}

\usepackage{tikz,pgfplots}
\usetikzlibrary{decorations.pathreplacing}
\usetikzlibrary{shapes.multipart}
\usetikzlibrary{positioning}
\usetikzlibrary{matrix}
\usetikzlibrary{shapes,calc,arrows}

\newcommand{\be}{\begin{equation}}
\newcommand{\ee}{\end{equation}}
\newcommand{\bea}{\begin{eqnarray}}
\newcommand{\eea}{\end{eqnarray}}

\title{Reflectionless pseudospin-1 Dirac systems via Darboux transformation and flat band solutions}

\author[${}$]{V. Jakubsk\'y${}^{1}$ and K. Zelaya${}^{1,2}$}

\affil[${1}$]{\footnotesize Nuclear Physics Institute, Czech Academy of Science, 250 68 \v{R}e\v{z}, Czech Republic}
\affil[${2}$]{\footnotesize Department of Physics, Queens College of the City University of New York, Queens, New York 11367, USA}

\date{}

\begin{document}
	
\maketitle
	
\begin{abstract}
This manuscript explores the Darboux transformation employed in the construction of exactly solvable models for pseudospin-one particles described by the Dirac-type equation. We focus on the settings where a flat band of zero energy is present in the spectrum of the initial system. Using the flat band state as one of the seed solutions substantially improves the applicability of the Darboux transformation, for it becomes necessary to ensure the Hermiticy of the new Hamiltonians. This is illustrated explicitly in four examples, where we show that the new Hamiltonians can describe quasi-particles in Lieb lattice with inhomogeneous hopping amplitudes.

\end{abstract}

\section{Introduction}
Darboux transformation is known for a long time in the analysis of differential equations \cite{Darboux}. It serves to map a differential equation into another one while keeping its solvability. In quantum physics, the transformation revived in the context of supersymmetric quantum mechanics. It forms there the super-charges and intertwining operators that provide a mapping between the superpartner Hamiltonians \cite{COOPER}. The transformation was used in the realm of condensed matter systems that are described by the low-dimensional Dirac equation. In that context, two different strategies were followed. The first one capitalized on the fact that the square of  a $2\times2$ one-dimensional Dirac operator with a minimally coupled magnetic field turns into a diagonal Schr\"odinger operator. Then, Darboux transformation for Sch\"odinger operators can be applied and the square root of the new Schr\"odinger operator is identified with the Dirac Hamiltonian of the new system, see e.g. \cite{KuruNegroNieto,MidyaFernandez,Jakubsky13} and more recent works \cite{Jahani,Phan20,CastilloCeleita20,Fernandez20,Fernandez20b,SchulzeRoy,Celeita,yesiltas,Concha-Sanchez,ContrerasNegro}.

In the second approach, Darboux transformation is applied directly to Dirac operators. This approach makes it possible to find solvable configurations of effective interactions that differ from magnetic fields. It was discussed for stationary one-dimensional, $2\times2$ Dirac equation in  \cite{Samsonov} and for non-stationary equation in \cite{Samsonov2}. Diverse extensions of Darboux transformation for Dirac operators were proposed for Dirac operators in polar coordinates~\cite{Sch22}, higher dimensions \cite{Schulze1,Iof22}, and/or higher spin~\cite{Schulze13}. A further generalization of the intertwining relations for Dirac operators was discussed in~\cite{Ioffe}.

The Darboux transformation of Dirac operators proved to be a great tool for analyzing confinement and scattering in diverse physical systems. For instance, it was used for the construction of solvable models of radially twisted carbon nanotubes \cite{Jakubsky12}. It was employed in the analysis of $PT$-symmetric optical settings described by $2\times2$ Dirac equation \cite{Correa17}. When applied to a free-particle system, it can produce a model with fluctuating potential that inherits trivial scattering characteristics of the free particle. In particular, Dirac fermion can tunnel the barrier without being backscattered. Such systems are called reflectionless. A class of reflectionless systems described by the one-dimensional Dirac equation was derived in  \cite{Correa14}, where their relation to the theory of integrable systems was discussed. An extended version of Darboux transformation was used to demonstrate the existence of omnidirectional (super) Klein tunneling of Dirac fermion in graphene through fluctuating 2D electrostatic barrier  \cite{Contreras20}.

Although the Darboux transformation is a powerful tool for the construction of solvable models, it is challenging to keep control over the form of interaction in the new Hamiltonian. Indeed, the new potential can fail to be hermitian. Additionally, not all of its entries can be associated with physical interaction.  Therefore, establishing the hermiticity of the potential term and its identification with actual physical interaction are important ingredients for implementing Darboux transformation in constructing physically meaningful models. This issue gets more challenging when the matrix coefficients of the Dirac operator are of higher dimensionality. For instance, the $4\times4$ equation is needed for the description of distortion scattering or spin-orbit interaction in graphene. Not all the matrix components of the potential produced by Darboux transformation can be associated with physical interactions. The problem was addressed for $4\times4$ Dirac operators in \cite{Celeita22}, where the form-preserving Darboux transformations were discussed for the class of reducible Dirac operators \cite{Celeita21}. 
 
Advances in experimental techniques allow for creation of artificial crystalline materials where collective excitations on the lattice behave like relativistic particles with either semi-integer or integer spin \cite{Ley18,Fan22}. Properties of these materials, e.g. geometry of the crystals, are highly tunable. The spin-one Dirac fermions emerge on the Lieb \cite{Gol11}, Kagome \cite{Mek03}, Dice \cite{Bercioux} or $\alpha-T_3$ lattices \cite{Raoux}. These artificial crystals can be composed with the use of optical lattices.  
The energy spectrum of spin-one particles can differ substantially from that of Dirac fermions in graphene. Additionally the Dirac cones, there can be a flat band of either zero energy (Lieb and Dice lattices) or of a non-vanishing energy (Kagome lattice). Further presence of flat bands has also been reported in slightly twisted bilayer graphene lattices~\cite{Mor10,Sin21}. This peculiar flat band represents a new opportunity in the effective use of Darboux transformation, which we aim to explore further in this article. 

This manuscript is organized as follows. First, the main features of Darboux transformation for Dirac-type operators are summarized. Then we briefly review emergence of spin-one Dirac equation from the tight-binding Hamiltonian of the Lieb lattice. We show how the effective interaction can be associated with peculiar properties of the lattice. In third section, we focus on the peculiarities of the use of Darboux transformation on the spin-one Dirac operators. We show on explicit examples that the new possibilities are available due to presence of the flat-band states. This way, a number of reflectionless models of spin-one particles are derived. The last section is left for discussion. 

\section{Darboux transformation for spin-one Dirac-type operators}
\label{sec:Darboux}
For completeness, we provide a brief overview of the Darboux transformation as applied to Dirac operators. Although this transformation can be used for systems with any (pseudo)spin, we focus on (pseudo)spin-1 systems. Let us consider an initial Hamiltonian $H$ and its corresponding eigenvalue equation
\begin{equation}
H=-i\hbar v_{f}\gamma\partial_{x}+V(x), \quad
H\boldsymbol{\Psi}_{E}=E\boldsymbol{\Psi}_{E}, \quad \partial_{x}\equiv\frac{\partial}{\partial x},
\end{equation}
respectively, where $\gamma$ and $V(x)$ are both Hermitian matrices in $\mathbb{C}^{3}$ so that $H$ renders a Hermitian operator in $L^{2}\otimes\mathbb{C}^{3}$. Thus, the eigenvalues $E$ are guaranteed to be real as long as their eigensolutions $\boldsymbol{\Psi}_{E}\in L^{2}\otimes\mathbb{C}^{3}$. 

We further assume that $H$ is exactly solvable so that its eigensolutions and eigenvalues are known. The goal is to construct a new unknown Hamiltonian $\widetilde{H}$ whose spectral information can be determined from that of $H$. Both Hamiltonians have the same kinetic term but different potentials. We thus have the explicit form of $\widetilde{H}$ and its eigenvalue problem
\begin{equation}
\widetilde{H}=-i\hbar v_{f}\gamma\partial_{x}+\tilde{V}(x), \quad 
\widetilde{H}\widetilde{\boldsymbol{\Psi}}_{{E}}={E}\widetilde{\boldsymbol{\Psi}}_{{E}} ,
\end{equation}
where the still unknown potential $\widetilde{V}(x)\in\mathbb{C}^{3\times 3}$ is determined from the Darboux transformation \cite{Samsonov}. The transformation is defined as 
\begin{equation}
\label{L}
L=\partial_{x}-U_{x}U^{-1},\quad HU=U\Lambda\end{equation} 
with $\Lambda$ being a fixed constant $3\times3$ matrix. When $\Lambda=diag(\epsilon_{1},\epsilon_{2},\epsilon_{3})$, the matrix $U=(\boldsymbol{\Psi}_{\epsilon_{1}},\boldsymbol{\Psi}_{\epsilon_{2}},\boldsymbol{\Psi}_{\epsilon_{3}})$ is constructed from three eigensolutions of $H$ associated with the eigenvalues $\epsilon_{j}$, with $j=1,2,3$.  
There holds 
\begin{equation}
\label{intertwining}
LH=\widetilde{H}L,
\end{equation}
provided that the potential $\widetilde{V}$ has the following form,
\begin{equation}
\widetilde{V}=V+i\hbar v_{f} [U_{x}U^{-1},\gamma] , 
\label{V-tilde}
\end{equation}
In order to have a well-defined potential $\widetilde{V}(x)$, one must impose the constraint $det(U)\neq 0$ for all $x\in Dom(V)$, so that $U^{-1}$ is well-defined and $\widetilde{V}(x)$  non-singular.

From the intertwining relations $LH=\widetilde{H}L$, it is clear that the $L$ maps eigensolutions of $H$ into the corresponding ones of $\widetilde{H}$, associated with the same eigenvalue $E$. That is, we have the mapping $\widetilde{\boldsymbol{\Psi}}_{E}\propto L\boldsymbol{\Psi}_{E} $. Furthermore, in analogy to the non-relativistic case, one has a set of eigensolutions that cannot be mapped through $L$, usually called \textit{missing state solutions} $\widetilde{\boldsymbol{\Psi}}_{\epsilon_{j}}$, with $j=1,2,3$. Such missing states are computed as the zero modes of $L^{\dagger}$, leading to the relation
\begin{equation}
\left( U^{-1} \right)^{\dagger}\equiv (\widetilde{\boldsymbol{\Psi}}_{\epsilon_{1}},\widetilde{\boldsymbol{\Psi}}_{\epsilon_{2}},\widetilde{\boldsymbol{\Psi}}_{\epsilon_{3}}), \quad \widetilde{H}\widetilde{\Psi}_{\epsilon_j}=\epsilon_j\widetilde{\Psi}_{\epsilon_j},\quad j=1,2,3.
\label{mapping-L}
\end{equation}
The new missing states $\boldsymbol{\Psi}_{\epsilon_{j}}$ might have a finite-norm even if the initial model does not admit finite-norm solutions for the same eigenvalues $\epsilon_{j}$. However, the latter is studied in a case-by-case scenario, for it depends on the specific form of the seed matrix $U$.
	

\section{Lieb Hamiltonian and pseudospin-1 Dirac equation}
\label{sec:Lieb}

\begin{figure}
	\centering
	\includegraphics[width=0.65\textwidth]{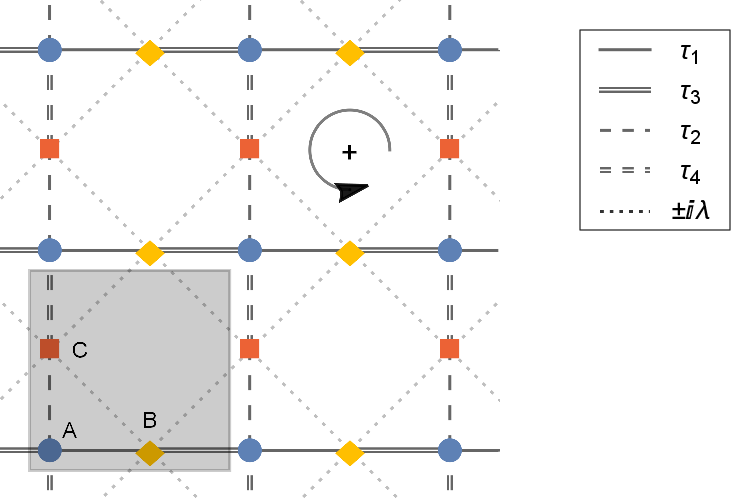}
    \caption{Lieb lattice with structure composed of nearest neighbor asymmetric shopping $\tau_{1}$ (solid line) and $\tau_{3}$ (double-line) between the atoms $A$ and $B$ in the $\hat{x}$-direction, as well as the hopping $\tau_{2}$ (dashed) and $\tau_{4}$ (double-dashed) for the atoms A and C along the $\hat{y}$-direction. Atoms $B$ and $C$ have a Haldane-like next-nearest neighbor interaction (dotted) denoted by $\pm i\lambda$. The sign is positive if the hooping happens counter-clockwise and negative otherwise. The shaded area depicts a unit cell.}
   \label{fig:Lieb}
\end{figure}

The Lieb lattice is  characterized by a two-dimensional line-centered array with three atoms per elementary cell (see the shaded area in Fig.~\ref{fig:Lieb}). 
The dynamics of the collective excitation under low-energy configuration can be approximated through  tight-binding model described by the following Hamiltonian:
\begin{equation}\label{TB-1}
H_{TB}=H_{NN}+H_{NNN}+H_{POT},
\end{equation}
where the total Hamiltonian has been split into the interactions
\begin{align}
H_{NN}&=-\sum_{\mathbf{R}_{A}}\sum_{j=1}^4\tau_j\mathcal{C}^{\dagger}_{\mathbf{R}_A+\delta_j}\mathcal{C}_{\mathbf{R}_A} +h.c. ,\\
H_{POT}&=\sum_{X=A,B,C}\mu_{X}\sum_{\mathbf{R}_X}\mathcal{C}^{\dagger}_{\mathbf{R}_X}\mathcal{C}_{\mathbf{R}_X} , \\
H_{NNN}&= - \sum_{\epsilon_1,\epsilon_2=\pm 1}\sum_{\mathbf{R}_C}t_{3}e^{i\mu \lambda}\mathcal{C}^{\dagger}_{\mathbf{R}_C+\epsilon_1\delta_1+\epsilon_2\delta_2}\mathcal{C}_{\mathbf{R}_C} +h.c. 
\end{align}
The position vectors $\mathbf{R}_{X}$, $X\in\{A,B,C\}$ runs over each of the three rectangular sublattices $A$, $B$, $C$ that form the Lieb lattice. The operator $\mathcal{C}^\dagger_{\mathbf{R}_{X}}$ creates the quasi-particle on the site $\mathbf{R}_{X}$ while $\mathcal{C}_{\mathbf{R}_{X}}$ annihilates the quasi-particle on this site.
 
The operator $H_{NN}$ represents anisotropic nearest-neighbor interaction. The nearest neighbor hopping parameters  $\tau_1$, $\tau_2$, $\tau_3$ and $\tau_4$ are real and can acquire different values. Such anisotropy appeared in \cite{Julku}, where superfluid characteristics of the flat band were discussed. Anisotropic hopping amplitudes were realized experimentally in \cite{Guzman-Silva} due to the asymmetric section of optical fibers that assembled the Lieb lattice. The on-site interaction $H_{POT}$ corresponds to the potential energy that can acquire different values $\mu_X$, $X\in\{A,B,C\}$ on the three sublattices $A,\ B,$ and $C$. The last term, $H_{NNN}$, stands for complex next-nearest neighbor interaction (NNN) between the atoms $B$ and $C$, which may be in general a complex-valued and direction sensitive quantity . Let us notice that such an interaction can  emerge due to external time-dependent fields in photonic Lieb lattices~\cite{Lon17}, and in magnon Lieb and Kagome lattices~\cite{Owe18}. Henceforth, we set $\lambda=\pi/2$ so that the NNN interaction resembles that introduced by Haldane~\cite{Hal88} as a prime example of anomalous quantum anomalous Hall effect in graphene. Such a term is also known in the literature as intrinsic spin-orbit coupling and has been exploited in other square lattices~\cite{Bha19}. Indeed, it was found experimentally in~\cite{Cha13}, whereas Landau levels in Lieb lattice with this NNN interaction were computed in \cite{Jak23a}. Such a NNN term was also implemented in dice lattices~\cite{Dey20} and in a honeycomb magnon lattice~\cite{Bos23} to investigate the topological properties and phase transitions. See also~\cite{Xin23} for a recent review.

The geometry of the Lieb lattice is shown in Fig. \ref{fig:Lieb},  where we depict a regular rectangular lattice with $\delta_3\equiv -\delta_1$ and $\delta_4\equiv -\delta_2$, $\delta_1=a(1,0)$, $\delta_2=a(0,1)$, $k=(k_x,k_y)$, and $a$ the atomic distance between nearest neighbors. As customary, we make a Fourier transform of the Hamiltonian $H_{TB}$ that in the momentum representation reads as 
\begin{equation}
H_{\operatorname{TB}}(\mathbf{k})=
-\begin{pmatrix}
-\mu_A & \tau_{1}e^{iak_{x}}+\tau_{3}e^{-iak_{x}} & \tau_{2}e^{iak_{y}}+\tau_{4}e^{-iak_{y}} \\
\tau_{1}e^{-iak_{x}}+\tau_{3}e^{iak_{x}} & -\mu_B & 4i\,t_3 \sin a k_x \sin a k_y \\
\tau_{2}e^{-iak_{y}}+\tau_{4}e^{iak_{y}} & -4i\, t_3 \sin a k_x \sin a k_y &- \mu_C
\end{pmatrix}
\end{equation}
from which the dispersion relation $E\equiv E(k)$ is obtained from the secular equation $\det (H(k)-E)=0$. This is an analytically solvable and third-order equation in $E$. These solutions define the energy bands whose structure can be quite rich in dependence on the parameters. 

For the purposes of the current article, we focus on the situation where a flat band of zero energy is present between two dispersion bands. Particularly, this occurs for $\tau_1=\tau_3$, $\tau_2=\tau_4$, and $\mu_j=0$, for $j=1,2,3,4$, leading to the dispersion relations
\begin{equation}
\label{dispspec1}
E_0(\mathbf{k})=0,\quad E_{\pm}(\mathbf{k})=\pm 2\sqrt{\tau_{1}^{2}\cos^{2}(ak_{x})+\tau_{2}^{2}\cos^{2}(ak_{y})+4t_{3}^{2}\sin^{2}(ak_{x})\sin^{2}(ak_{y})}.
\end{equation}
Alternatively, the parameters $\mu_j=0$ and $t_3=0$ also support a flat band so that the dispersion relations become
\begin{equation}
\label{dispspec2}
E_0(\mathbf{k})=0,\quad E_{\pm}(\mathbf{k})=\pm\sqrt{\sum_{j=1}^4\tau_j^2+2\tau_1\tau_3\cos 2k_x+2\tau_2\tau_4\cos 2k_y}.
\end{equation}

The dispersion relations~\eqref{dispspec1}-\eqref{dispspec2} show that a band gap emerges whenever anisotropic hopping amplitudes or non-vanishing NNN interactions are considered. In both cases, the band gap emerges around the Dirac valley (local minimum) located at $\mathbf{K}_0=\{\pi/2,\pi/2\}$ in the primi\-tive unitary cell. The minimum deviates from $\mathbf{K}_0$ for generic values of $\tau_{1,2,3,4}$, $t_3$ and $\mu_{1,2,3}$. Despite the latter, by expanding the tight-binding Hamiltonian $H_{\operatorname{TB}}\left(\mathbf{k}\right)$ in the vicinity of $\mathbf{K}_0$, $\mathbf{k}=\mathbf{K}_0+\boldsymbol{\delta k}$, with $\Vert\mathbf{\delta k}\Vert\ll 1$, we can provide an accurate approximation for the electron dynamics provided that the minimum is close to $\mathbf{K}_0$. From this consideration, the expanded Hamiltonian takes the form
\begin{equation}
 H^{\operatorname{ex}}_{\operatorname{TB}}\approx a(\tau_1+\tau_3)S_1\delta k_x+a(\tau_2+\tau_4)S_2\delta k_y +\operatorname{diag}\{\mu_1,\mu_2,\mu_3\}+(\tau_1-\tau_3)\widetilde{S}_1+(\tau_2-\tau_4)\widetilde{S}_2+4 t_3 S_3 ,
\label{HDP}
\end{equation}
where we have introduced the matrices
\begin{equation}
\begin{aligned}
& S_{1}=
\begin{pmatrix}
0 & 1 & 0 \\
1 & 0 & 0 \\
0 & 0 & 0
\end{pmatrix}
,\quad S_{2}=
\begin{pmatrix}
0  & 0 & 1 \\
0 & 0 & 0 \\
1  & 0 & 0
\end{pmatrix},  \quad
S_{3}=
\begin{pmatrix}
0 & 0 & 0 \\
0 & 0 & -i \\
0 & i & 0
\end{pmatrix}
, 
\\
& \widetilde{S}_{1}=
\begin{pmatrix}
0  & -i & 0 \\
i & 0 & 0 \\
0  & 0 & 0
\end{pmatrix} 
, \quad
\widetilde{S}_{2}=
\begin{pmatrix}
0  & 0 & -i \\
0  & 0 & 0  \\
i  & 0 & 0
\end{pmatrix}
.
\end{aligned}
\label{matrix-spin1}
\end{equation}
In turn, the following matrices are useful for the upcoming discussion:
\begin{equation}
S=\begin{pmatrix}
1 & 0 & 0 \\
0 & -1 & 0 \\
0 & 0 & 1 
\end{pmatrix}
,
\quad
\widetilde{S}_{3}=
\begin{pmatrix}
1  & 0  & 0 \\
0  & -1 & 0  \\
0  & 0  & 0
\end{pmatrix}
.
\label{matrix-S3-S3t}
\end{equation}

Although it is possible to find the eigenstates of the Hamiltonian~\eqref{HDP}, their explicit form is rather complicated. We instead focus on the two specific situations discussed in~\eqref{dispspec1}-\eqref{dispspec2} where the flat band exists. We will be interested in the situations where the fluctuation of the coupling parameters $\tau_{1,2,3,4}$, $\mu_{1,2,3}$, or $t_3$ can localize the quasi-particles.


\section{Darboux-transformed reflectionless systems}
\label{sec:Darboux-Lieb}
In this section, we aim to construct exactly solvable models described by~\eqref{HDP} with possibly inhomogeneous interaction caused by varying hopping amplitudes. The models discussed here are generated through the Darboux transformation applied on the free-particle pseudospin-one Hamiltonian. The spectrum of the resulting model is that of free particles with possibly the addition of up to three bound states eigenvalues.

Particularly, we depart from the free-particle model of the form (see~\cite{Jak23b} for details)
\begin{equation}
H(x)\boldsymbol{\Psi}(x) \equiv ( -i\hbar v_{f}S_{1}\partial_{x} + m S_{3} )\boldsymbol{\Psi}(x)=E\boldsymbol{\Psi}(x) ,
\label{Dirac-H2}
\end{equation}
with $m$ the mass-term and  $\boldsymbol{\Psi}(x)=(\psi_{A}(x),\psi_{B}(x),\psi_{C}(x))^{T}$ the eigenfunction associated with the eigenvalue $E$. This particular free particle model possesses a chiral symmetry generated by $S$, i.e., $\{S,H\}=0$. Thus, if $\boldsymbol{\Psi}_{\epsilon}$ is an eigensolution of $H$ for $E=\epsilon$, then $\boldsymbol{\Psi}_{-\epsilon}=S\boldsymbol{\Psi}_{\epsilon}$ solves $H$ for $E=-\epsilon$. Such symmetry plays a fundamental role in the construction of new models.

Considering the explicit solutions of (\ref{Dirac-H2}), it is customary to consider the cases $|\epsilon|\neq |m|$ and $\epsilon=\pm m$ separately. In the first case, the equation (\ref{Dirac-H2}) can be rewritten into decoupled form
\begin{align}
\hbar^{2}v_{f}^{2}\psi_{A}''=(m^{2}-\epsilon^{2})\psi_{A} , \quad
\psi_{B}=i\frac{\hbar v_{f} \epsilon \psi_{A}'}{m^{2}-\epsilon^{2}} , \quad 
\psi_{C}=-\frac{\hbar m v_{f} \psi_{A}'}{m^{2}-\epsilon^{2}} , \quad \vert\epsilon\vert\neq \vert m\vert ,
\label{psiABC}
\end{align}
so that one obtains the solution $\boldsymbol{\Psi}_{\epsilon}=e^{\pm\nu x}\left( -(m^{2}-\epsilon^{2}) , \pm\hbar v_{f} \nu \epsilon , \pm i \hbar m v_{f}\nu \right)^{T}$, with $\hbar^{2}v_{f}^{2}\nu^{2}=m^{2}-\epsilon^{2}$. 
We can use the latter and make proper linear combinations to construct a more convenient set of solutions in terms of hyperbolic functions, as presented in Table~\ref{tab:solutions}. In turn, for $\epsilon=\pm m$, we get in particular that 
\begin{equation}
\psi_{A}=\ell\in\mathbb{C},\quad -i\hbar v_{f}\psi_{B}'=\pm m \ell,\quad \psi_{C}=\pm i \psi_{B},\quad \epsilon=\pm m ,
\end{equation}
the solutions of which are also presented in Table~\ref{tab:solutions}. 

The eigenvalue $\epsilon=0$ is specific. There are solutions obtained by performing the limit $\epsilon\rightarrow 0$ in (\ref{psiABC}). Additionally, there are \textit{flat-band solutions}. These are obtained by fixing $\epsilon=0$ in the Dirac equation~\eqref{Dirac-H2} and decoupling the equations. We obtain $\psi_{B}'=\psi_{B}=0$ and $m\psi_{C}=\hbar v_{f}\psi_{A}'$. Thus, there is no unique solution, and one can write the general solutions as 
\begin{equation*}
\boldsymbol{\Psi}_{fb}=(m\chi,0,-\hbar v_{f}\chi')^{T},
\end{equation*} 
where $\chi$ is an arbitrary complex-valued function. The arbitrariness of $\chi(x)$ can be  understood by considering the particular flat-band solution $\boldsymbol{\Psi}_{fb;\kappa}=e^{i\kappa x}(m,0,-i\kappa\hbar v_{f})$, i.e.  $\chi_{\kappa}(x)=e^{i\kappa x}$, where $\kappa\in\mathbb{R}$. Such solutions are called degenerated plane waves (or \textit{degenerate Bloch wave}~\cite{Ley18}). One can make use of these degenerate plane waves to compose an arbitrary wave packet through the linear combination $\boldsymbol{\Psi}_{fb}(x)=\int_{\mathbb{R}}d\kappa f(\kappa)\boldsymbol{\Psi}_{\kappa;fb}(x)$ with $f(\kappa)$ a complex-valued function. Two specific choices of $\chi$ and the generic one are presented in the last line of Table~\ref{tab:solutions}.

\begin{table}
	\centering
	\begin{tabular}{c|c|l}
		Energy $\epsilon$ & $\boldsymbol{\Psi}_{\epsilon}(x)$ & Remarks \\ \hline
		\multirow{2}{*}{$-m<\epsilon<m$} & $\left( \hbar v_{f}\nu \sinh(\nu x) , i\epsilon \cosh(\nu x) , -m \cosh(\nu x) \right)^{T}$ & \multirow{2}{*}{$\nu=\frac{\sqrt{m^{2}-\epsilon^{2}}}{\hbar v_{f}}$} \\
		& $\left( \hbar v_{f}\nu \cosh(\nu x) , i\epsilon \sinh(\nu x) , -m \sinh(\nu x) \right)^{T}$ &  \\
		\hline
		\multirow{2}{*}{$\pm m$} & \multirow{2}{*}{$(\ell_{1} \hbar v_{f}  , \pm i \ell_{1} m x+\ell_{0}, -\ell_{1} m x \pm i\ell_{0})^{T}$} & \multirow{2}{*}{$\ell_{0},\ell_{1}\in\mathbb{C}$} \\
		&  &  \\
		\hline
		\multirow{4}{*}{$0$} & $\left( \hbar v_{f}\nu_{0} \sinh(\nu_{0} x) , 0 , -m \cosh(\nu_{0} x) \right)^{T}$ & \multirow{2}{*}{$\nu_{0}=\frac{m}{\hbar v_{f}}$} \\
		& $\left( \hbar v_{f}\nu_{0} \cosh(\nu_{0} x) , 0 , -m \sinh(\nu_{0} x) \right)^{T}$ &  \\
		& \multirow{2}{*}{$\left( m\chi(x),0,-\hbar v_{f}\chi'(x) \right)^{T}$}   & \multirow{2}{*}{Flat band solution,} \\
		&  & $\chi(x)$ arbitrary
	\end{tabular}
	\caption{Free particle solutions for the Lieb lattice at three different energy regimes, including the degenerate solutions at the flat band energy.}
	\label{tab:solutions}
\end{table}

In the rest of the section, we will use the system described by (\ref{Dirac-H2}) as the initial model for Darboux transformation. In accordance with Sec.~\ref{sec:Lieb}, the transformation $L$ as well as the new potential $\widetilde{V}(x)$ are defined in terms of three eigenvectors of $H$ that correspond to the factorization energies $\epsilon_{1}$, $\epsilon_{2}$ and $\epsilon_{3}$. We define $\overline{V}(x)$ as the difference between the initial and the new potential,
\begin{equation}
\widetilde{V}(x)=m S_{3}+\overline{V}(x), \quad \overline{V}=i \hbar v_{f} [U_{x}U^{-1},S_{1}], 
\label{V-tilde-2}
\end{equation}
where the matrix $U$ is defined below (\ref{L}). 

In the next section, we present four particular models that illustrate how using a flat band solution makes it possible to keep the new operator Hermitian. Additionally, we show that the Darboux transformation leads to new non-trivial results even when two factorization energies coincide. This is in stark contrast with the pseudospin-$1/2$ case, where such factorization energies lead to new models that coincide with the initial one.


\subsection{Case $\Lambda=diag(\epsilon,\epsilon=0,-\epsilon)$}
\label{CaseI}
In order to illustrate the usefulness of the flat band solution in the Darboux transformation, let us consider the 
factorization energies as 
\begin{equation}
\epsilon_1=\epsilon,\ \epsilon_2=0,\ \epsilon_3=-\epsilon,\quad |\epsilon|<m.
\label{epsilon-case-1}
\end{equation}
There are different ways to combine the seed solutions $\boldsymbol{\Psi}_{\epsilon}(x)$. First, we focus on using a zero-energy solution $\boldsymbol{\Psi}_{\epsilon=0}(x)$ instead of the flat band solution $\boldsymbol{\Psi}_{fb}(x)$ for $\epsilon=0$. As we shall see, avoiding the flat band solutions results in manifestly non-hermitian Hamiltonian. We fix the seed matrix as
\begin{equation}
U=\left( \boldsymbol{\Psi}_{\epsilon}, \boldsymbol{\Psi}_{\epsilon=0}, \boldsymbol{\Psi}_{-\epsilon}\right)=
\begin{pmatrix}
\sinh(\nu x) & \cosh(\nu_{0}x) & \sinh(\nu x) \\
i\frac{\epsilon}{\hbar v_{f}\nu}\cosh(\nu x) & 0 & -i\frac{\epsilon}{\hbar v_{f}\nu}\cosh(\nu x) \\
-\frac{m}{\hbar v_{f} \nu}\cosh(\nu x) & - \frac{m}{\hbar v_{f} \nu_{0}} \sinh(\nu_{0}x) & -\frac{m}{\hbar v_{f}\nu}\cosh(\nu x)
\end{pmatrix}
,
\label{U-11}
\end{equation}
where $\nu=\sqrt{m^{2}-\epsilon^{2}}/\hbar v_{f}$ and $\nu_{0}=m/\hbar v_{f}$, and its determinant takes the form
\begin{equation}
\Delta(x):=\operatorname{det}(U(x))=\frac{2i m \epsilon}{\hbar^{2} v_{f}^{2}\nu\nu_{0}} \cosh^{2}(\nu x)\cosh(\nu_{0}x) \left( -\frac{\nu_{0}}{\nu}+\tanh(\nu x)\tanh(\nu_{0}x) \right) ,
\label{detU-11}
\end{equation}
which is non-null provided that $\nu_{0}/\nu>1$, a property guaranteed for $-m<\epsilon<m$. 

The straightforward calculations show that the new potential $\overline{V}(x)$ in~\eqref{V-tilde-2} has the null components $\overline{V}_{11}=\overline{V}_{22}=\overline{V}_{33}=\overline{V}_{13}=\overline{V}_{31}=0$, whereas the non-vanishing components are explicitly given by
\begin{equation}
\begin{aligned}
& \overline{V}_{12}=V_{21}^{*}=-i\frac{\psi_{B;\epsilon}'}{\psi_{B;\epsilon}}+i\frac{\psi_{C;\epsilon}\psi_{A;0}'-\psi_{A;\epsilon}'\psi_{C;0}}{\widetilde{\Delta}} , \\
& \overline{V}_{23}=\frac{2i\psi_{B;\epsilon}\left( \psi_{A;\epsilon}'\psi_{A;0}-\psi_{A;\epsilon}\psi_{A;0}' \right)}{\Delta}=\frac{i\hbar v_{f}\nu_{0}\nu}{m}\left(\frac{-\nu+\nu_{0}\tanh(\nu_{0}x)\tanh(\nu x)}{-\nu_{0}+\nu\tanh(\nu_{0}x)\tanh(\nu x)} \right), \\ 
& \overline{V}_{32}=\frac{2i\psi_{B;\epsilon}\left(\psi_{C;\epsilon}'\psi_{C;0}-\psi_{C;\epsilon}\psi_{C;0}' \right)}{\Delta}=\frac{im}{\hbar v_{f}},
\end{aligned}
\label{case-1-V23}
\end{equation}
where we have used the shorthand notation $\boldsymbol{\Psi}_{\epsilon}=(\psi_{A;\epsilon},\psi_{B;\epsilon},\psi_{C;\epsilon})^{T}$ and $\boldsymbol{\Psi}_{\epsilon=0}=(\psi_{A;0},0,\psi_{C;0})^{T}$, with the components taken from~\eqref{U-11}. 

We now impose the necessary conditions on the available parameters so that $\widetilde{H}$ becomes a Hermitian operator in $L^{2}\otimes\mathbb{C}^{3}$, which is achieved whenever $\overline{V}_{23}=\overline{V}_{32}^{*}$. The latter is held if and only if $m=\epsilon=0$, which contradicts our initial condition of a non-mull mass term. It is then concluded that $\widetilde{H}$ cannot be Hermitian in $L^{2}\otimes\mathbb{C}^{3}$ if the seed matrix~\eqref{detU-11} is used. 

Now, let us show that one can still use the factorization energies in~\eqref{epsilon-case-1} and achieve the desired Hermiticity provided that flat band solution is employed. To this end, let us make an alternative choice of the seed solutions by fixing 
the seed matrix as 
\begin{equation}
U(x)=\left( \boldsymbol{\Psi}_{\epsilon}, \boldsymbol{\Psi}_{fb}, \boldsymbol{\Psi}_{-\epsilon}\right)=
\begin{pmatrix}
\sinh(\nu x) & m \chi(x) & \sinh(\nu x) \\
i\frac{\epsilon}{\hbar v_{f}\nu}\cosh(\nu x) & 0 & -i\frac{\epsilon}{\hbar v_{f}\nu}\cosh(\nu x) \\
-\frac{m}{\hbar v_{f} \nu}\cosh(\nu x) & -\hbar v_{f} \chi'(x) & -\frac{m}{\hbar v_{f}\nu}\cosh(\nu x)
\end{pmatrix}
,
\label{case-1-U}
\end{equation}
where $\chi(x)$ is an unknown and arbitrary function to be specified later and related to the degene\-rate Bloch functions.

Similar to the previous case, the potential term $\overline{V}(x)$ has the same null components, whereas the only non-vanishing components are $\overline{V}_{12}$, $\overline{V}_{21}$, $\overline{V}_{23}$, and $\overline{V}_{32}$. Furthermore, it follows that $\overline{V}_{12}=\overline{V}_{21}^{*}$ as long as $\chi(x):\mathbb{R} \mapsto\mathbb{R}$. The exact form of $\chi(x)$ is then determined by imposing $\overline{V}_{23}=\overline{V}_{32}^{*}$, the form of which can be determined from~\eqref{case-1-V23} by making the change $\psi_{A;0}\rightarrow\psi_{A;fb}-m\chi(x)$ and $\psi_{C;0}\rightarrow\psi_{C;fb}=-\hbar v_{f}\chi'(x)$.  After some calculations one arrives to the relation $\chi''(x)=\nu^{2}\chi(x)$, from which we choose the solution
\begin{equation}
\chi(x)=\cosh(\nu x) , \quad \boldsymbol{\Psi}_{fb}(x)=(m \cosh(\nu x),0,-\hbar v_{f}\nu\sinh(\nu x))^{T}.
\end{equation}

In this form, the determinant of the seed matrix becomes
\begin{equation}
\Delta(x):=\det(U(x))=-\frac{2i\epsilon}{\hbar v_{f}}\cosh^{3}(\nu x)\left( -\frac{m^{2}}{m^{2}-\epsilon^{2}} +\tanh^{2}(\nu x) \right) 
\end{equation}
which is non-null for all $x\in\mathbb{R}$, leading to an invertible seed matrix\footnote{The solution $\boldsymbol{\Psi}_{fb}(x)=(m \sinh(\nu x),0,-\hbar v_{f}\nu\cosh(\nu x))^{T}$ leads to a singular matrix $U(x)$ at $x=0$.} $U(x)$. 

\begin{figure}
    \centering
    \subfloat[][]{\includegraphics[width=0.4\textwidth]{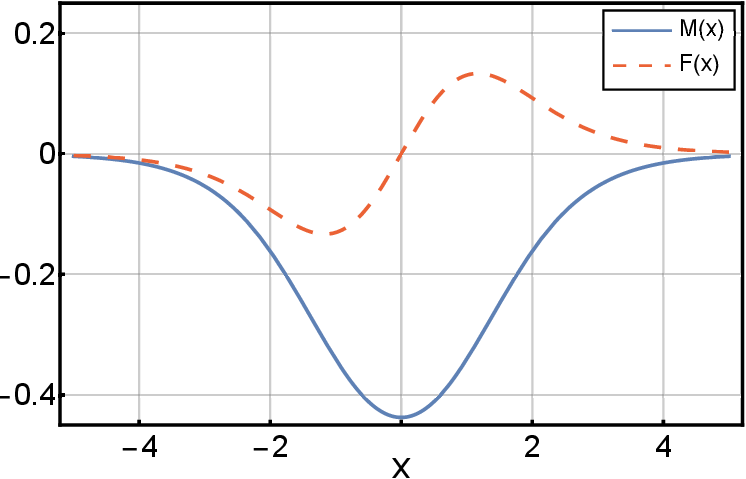}
    \label{fig:pot1a}}
    \hspace{2mm}
    \subfloat[][]{\includegraphics[width=0.4\textwidth]{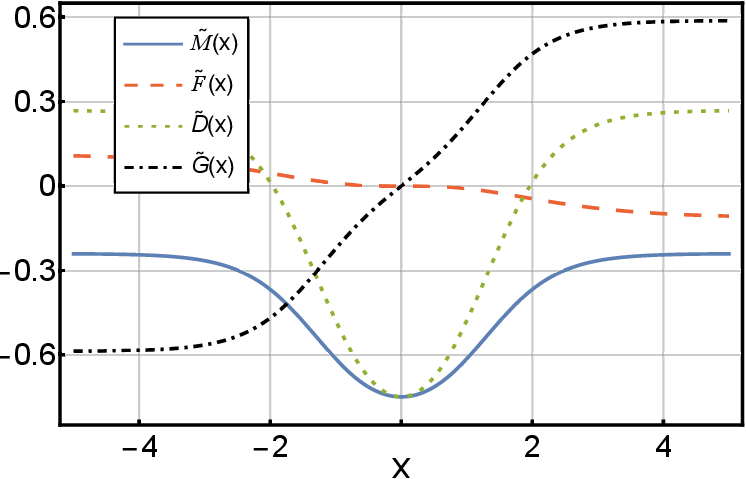}
    \label{fig:pot1b}}
    \caption{Components of the new matrix potential $\widetilde{V}(x)$ for the case $\Lambda=diag(\epsilon,0,-\epsilon)$ (a) and $\Lambda=diag(m,0,\epsilon)$ (b). Here, the set of parameters have been fixed as $\{\epsilon=0.75,\hbar=v_{f}=m=1\}$ (a) and $\{\epsilon=-0.25,\hbar=v_{f}=m=1\}$ (b).}
    \label{fig:pot1}
\end{figure}

From these consideration, the newly generated Hamiltonian $\widetilde{H}$ is regular, Hermitian in $L^{2}\otimes\mathbb{C}^{3}$, and takes the explicit form
\begin{equation}
\widetilde{H}=-i\hbar v_{f}S_{1}\partial_{x}+(m+M(x))S_{3}+F(x)\widetilde{S}_{1} , 
\end{equation}
with the inhomogeneous components
\begin{equation}
F(x)= -\frac{\hbar v_{f}\nu\tanh(\nu x)\operatorname{sech}^{2}(\nu x)}{-\frac{m^{2}}{\hbar^{2}v_{f}^{2}\nu^{2}}+\tanh^{2}(\nu x)} , \quad M(x)= \frac{m\operatorname{sech}^{2}(\nu x)}{-\frac{m^{2}}{\hbar^{2}v_{f}^{2}\nu^{2}}+\tanh^{2}(\nu x)} .
\end{equation}
that can be interpreted as asymptotically vanishing inhomogeneities of the NNN interaction and the hopping parameters $\tau_1$ and $\tau_3$ such that $\tau_1+\tau_3$ is constant. It is worth noting that $\{S,\widetilde{S}_{1}\}=0$ and thus $\widetilde{H}$ preserves the chiral symmetry of the initial system, $\{S,\widetilde{H}\}=0$. The potential term $\widetilde{V}(x)$ is asymptotically equal to that of the free particle, as illustrated in Fig.~\ref{fig:pot1a}. The functions $F(x)$ and $M(x)$ represent localized fluctuations in $x$ of the NN and NNN hopping amplitudes $\tau_1$ and $\tau_3$, compare with (\ref{HDP}). These fluctuations are responsible for the confining of quasi-particles. Indeed, the three missing state solutions~\eqref{mapping-L}, computed as the zero modes of $(U^{-1})^{\dagger}$, are square-integrable and given, up to a normalization constant, as
\begin{equation}
\widetilde{\boldsymbol{\Psi}}_{\epsilon}=\mathcal{N}_{\epsilon}\begin{pmatrix}
\frac{\tanh(\nu x)\operatorname{sech}(\nu x)}{-\frac{m^{2}}{\hbar^{2}v_{f}^{2}\nu^{2}}+\tanh^{2}(\nu x)} \\
i\frac{\hbar v_{f}\nu}{\epsilon}\operatorname{sech}(\nu x) \\
\tfrac{m}{\hbar v_{f}\nu}\frac{\operatorname{sech}(\nu x)}{-\frac{m^{2}}{\hbar^{2}v_{f}^{2}\nu^{2}}+\tanh^{2}(\nu x)}
\end{pmatrix}
, \quad
\widetilde{\boldsymbol{\Psi}}_{-\epsilon}=S\widetilde{\boldsymbol{\Psi}}_{\epsilon} , \quad 
\widetilde{\boldsymbol{\Psi}}_{0}=\mathcal{N}_{0}
\begin{pmatrix}
-\frac{\tfrac{m}{\hbar^{2}v_{f}^{2}\nu^{2}}\operatorname{sech}(\nu x)}{-\frac{m^{2}}{\hbar^{2}v_{f}^{2}\nu^{2}}+\tanh^{2}(\nu x)} \\
0 \\
- \frac{\tfrac{1}{\hbar v_{f}\nu}\tanh(\nu x)\operatorname{sech}(\nu x)}{-\frac{m^{2}}{\hbar^{2}v_{f}^{2}\nu^{2}}+\tanh^{2}(\nu x)}
\end{pmatrix}
.
\label{case-1-missing}
\end{equation}
These missing state solutions are the only bound states of $\widetilde{H}$, and the point spectrum of $\widetilde{H}$ becomes $\sigma(\widetilde{H})=\{\epsilon,0,-\epsilon\}$, with $0<\epsilon<m$. 

\subsection{Case $\Lambda=diag(m,0,\epsilon)$}
\label{caseII}
As a second example, we consider the alternative set of factorization energies
\begin{equation}
\epsilon_1=m,\quad \epsilon_2=0,\quad-m<\epsilon_3\equiv\epsilon<0,
\label{epsilon-case-2}
\end{equation}
where $\epsilon_{3}$ is considered negative to ensure the regularity of the transformation, as shown below.

Following the result of the previous case, we use the flat-band solutions for $\epsilon=0$ and take the solution at $\epsilon=m$ from Table~\ref{tab:solutions} with $\ell_{1}=0$ and $\ell_{0}=1$. The seed matrix becomes
\begin{equation}
U(x)=\left( \boldsymbol{\Psi}_{m}, \boldsymbol{\Psi}_{fb}, \boldsymbol{\Psi}_{\epsilon}\right)=
\begin{pmatrix}
0 & m \chi(x) & \sinh(\nu x) \\
1 & 0 & i\frac{\epsilon}{\hbar v_{f}\nu}\cosh(\nu x) \\
i & -\hbar v_{f} \chi'(x) & -\frac{m}{\hbar v_{f}\nu}\cosh(\nu x)
\end{pmatrix}
.
\label{case-2-U}
\end{equation}

For simplicity, the determinant of $U(x)$ will be discussed once we compute the components of $\widetilde{V}(x)$, which has $\overline{V}_{33}=0$ as the only vanishing component. We further constraint $\chi(x)$ as a real-valued function so that $\overline{V}_{12}=\overline{V}_{21}^{*}$ is automatically ensured and  $\overline{V}_{11}=-\overline{V}_{22}$, with $\overline{V}_{11,22}$ real-valued functions. We have to satisfy the following equations $\overline{V}_{13}=\overline{V}_{31}^{*}$ and $\overline{V}_{23}=\overline{V}_{23}^{*}$ so that $\widetilde{V}(x)$ becomes Hermitian. 

In this case, the determinant $\Delta(x):=\det(U(x))=\psi_{A;fb}\psi_{C;\epsilon}-\psi_{C;fb}\psi_{A;\epsilon}-i\psi_{B;\epsilon}\psi_{A;fb}$ becomes a real quantity. Thus, from the explicit form of the potential components 
\begin{equation}
\begin{alignedat}{3}
& \overline{V}_{13}(x)= \frac{m\epsilon\chi \sinh(\nu x)}{\hbar v_{f} \Delta(x)}, \quad && \overline{V}_{31}(x)= \frac{\sinh(\nu x)}{\hbar v_{f} \Delta(x)}(\hbar^{2}v_{f}^{2}\chi''-m^{2}\chi), \\
& \overline{V}_{23}=-i\frac{m\cosh(\nu x)}{\Delta(x)}(\nu \chi-\chi'\operatorname{tanh}(\nu x)) , \quad && \overline{V}_{32}=-i\frac{\cosh(\nu x)}{\nu \Delta(x)}\left( m\nu\chi'\operatorname{tanh}(\nu x) +(\epsilon-m)\chi'' \right) ,
\end{alignedat}
\end{equation}
one notices that $\overline{V}_{13}$ and $\overline{V}_{31}$ are real-valued functions, whereas $\overline{V}_{23}$ and $\overline{V}_{32}$ are imaginary functions. Therefore, one must impose the constraints $\overline{V}_{13}=\overline{V}_{31}$ and $\overline{V}_{23}^{*}=\overline{V}_{32}$ so that the resulting potential $\widetilde{V}(x)$ is Hermitian. Interestingly, we have two equations and only one free function $\chi(x)$. Still, it is straightforward to note that the condition $\chi''=\tfrac{m(m+\epsilon)}{\hbar^{2}v_{f}^{2}}\chi$ fulfills both constraints simultaneously.

From the latter, one thus chooses the flat-band solution as
\begin{equation}
\chi(x)=\cosh(\sigma x) , \quad \boldsymbol{\Psi}_{fb}(x)=(m\cosh(\sigma x)),0,-\hbar v_{f}\sigma \sinh(\sigma x)^{T}, \quad \sigma=\frac{\sqrt{m(m+\epsilon)}}{\hbar v_{f}} ,
\end{equation}
so that the determinant
\begin{equation}
\Delta(x)=\hbar v_{f}\sigma\cosh(\sigma x)\cosh(\nu x)\left(-\frac{\sqrt{m(m-\epsilon)}}{m+\epsilon}+\tanh(\sigma x)\tanh(\nu x) \right) 
\end{equation}
is non-singular for all $x\in\mathbb{R}$ if $-m<\epsilon<0 $, in accordance with the constraint introduced in~\eqref{epsilon-case-2}. Moreover, the new Hamiltonian $\widetilde{H}$ is Hermitian and takes the form
\begin{equation}
\widetilde{H}=-i\hbar v_{f}S_{1}\partial_{x}+(m+\widetilde{M}(x))S_{3}+\widetilde{G}(x)S_{2}+\widetilde{F}(x)\widetilde{S}_{1}+\widetilde{D}(x)\widetilde{S}_{3},
\end{equation}
where the inhomogeneous terms are
\begin{equation}
\begin{aligned}
& \widetilde{F}(x)= \frac{\epsilon\cosh(\sigma x)\cosh(\nu x)}{\Delta(x)}\left(-\sqrt{m(m-\epsilon)}\tanh(\sigma x)+m\tanh(\nu x) \right), \\
& \widetilde{M}(x)= \frac{m\sqrt{m(m+\epsilon)}\cosh(\sigma x)\cosh(\nu x)}{\Delta(x)}\left(\sqrt{\frac{m-\epsilon}{m}}-\tanh(\sigma x)\tanh(\nu x) \right) , \\
& \widetilde{D}(x)= \frac{(m-\epsilon)\sqrt{m(m+\epsilon)}\cosh(\sigma x)\cosh(\nu x)}{\Delta(x)}\left( \sqrt{\frac{m}{m-\epsilon}}-\tanh(\sigma x)\tanh(\nu x) \right) , \\
& \widetilde{G}(x)=\frac{m \epsilon \cosh(\sigma x)\sinh(\nu x)}{\Delta(x)} .
\end{aligned}
\end{equation}
The term $\widetilde{G}(x)$ associated with $S_2$ can be attributed to the vector potential, the term $\widetilde{M}S_3$ corresponds to an inhomogeneous NNN interaction, and $\widetilde{F}\widetilde{S}_1$ can be associated with inhomogeneous hopping amplitudes, see (\ref{HDP}). The last term $\widetilde{D}(x)S_3$ represents an onsite interaction. The behavior of these terms is depicted in Fig.~\ref{fig:pot1b}, where it is clear that the interactions are not asymptotically symmetric. This is indeed one of the main differences with respect to the case discussed in Sec.~\ref{CaseI}.
  
In this model, only the missing states corresponding to the eigenvalues $E=0$ and $E=\epsilon$ are square-integrable. Their explicit forms are
\begin{equation}
\widetilde{\boldsymbol{\Psi}}_{0}(x)=
\begin{pmatrix}
& -\sqrt{\frac{m-\epsilon}{m+\epsilon}}\frac{\cosh(\nu x)}{\Delta(x)} \\
& -\frac{i\sinh(\nu x)}{\Delta(x)} \\
& -\frac{\sinh(\nu x)}{\Delta(x)}
\end{pmatrix}
, \quad
\widetilde{\boldsymbol{\Psi}}_{\epsilon}(x)=
\begin{pmatrix}
& \frac{\sqrt{m(m+\epsilon)}\sinh(\sigma x)}{\Delta(x)} \\
& i\frac{m\cosh(\sigma x)}{\Delta(x)} \\
& \frac{m \cosh(\sigma x)}{\Delta(x)}
\end{pmatrix}
,
\end{equation}
and the point spectrum is composed of only two eigenvalues, namely, $\sigma(\widetilde{H})=\{0,\epsilon\}$, with $-m<\epsilon<0$. It is worth noting that this case can also be presented by taking into account the factorization energies $\epsilon=-m$, $\epsilon_{2}=0$, and $0<\epsilon_{3}=\epsilon<m$. By doing so, analog calculations follow and result in the point spectrum $\sigma(\widetilde{H})=\{0,\epsilon\}$, with $0<\epsilon<m$.

\subsection{Case $\Lambda=diag(\epsilon,\epsilon,0)$}
\label{caseIII}
Now, let us demonstrate that nontrivial results can be obtained despite the fact that two of the three factorization energies coincide. To this end, we fix
\begin{equation}
\epsilon_1=\epsilon_2=\epsilon,\quad\epsilon_3=0,\quad |\epsilon|<|m|,
\end{equation}
and use the two linearly-independent solutions $\boldsymbol{\Psi}_{\epsilon_1}$ and $\boldsymbol{\Psi}_{\epsilon_2}$ for $\epsilon_1=\epsilon_2=\epsilon$ shown in Table~\ref{tab:solutions}. The third selected eigenvector is an arbitrary flat-band solution $\boldsymbol{\Psi}_{fb}$, which defines the seed matrix 
\begin{equation}
U(x)=\left( \boldsymbol{\Psi}_{\epsilon_1}, \boldsymbol{\Psi}_{\epsilon_2}, \boldsymbol{\Psi}_{fb}\right)=
\begin{pmatrix}
\sinh(\nu x) & \cosh(\nu x) & m \chi(x)  \\
i\frac{\epsilon}{\hbar v_{f}\nu}\cosh(\nu x) & i\frac{\epsilon}{\hbar v_{f}\nu}\sinh(\nu x) & 0 \\
-\frac{m}{\hbar v_{f}\nu}\cosh(\nu x) & -\frac{m}{\hbar v_{f}\nu}\sinh(\nu x) & -\hbar v_{f} \chi'(x)
\end{pmatrix}
.
\label{case4-U}
\end{equation}
The determinant reads as $\Delta(x)=\det(U(x))=i\epsilon\chi'(x)/\nu$ so that $\chi(x)$ shall be a monotonous function in order to ensure the invertibility of $U(x)$.

The direct computation of the new potential components reveal that  $\overline{V}_{11}(x)=\overline{V}_{22}(x)=\overline{V}_{33}(x)=0$. Next, for a real-valued $\chi(x)$, one obtains $\overline{V}_{12}(x)=\overline{V}_{12}^{*}(x)=\tfrac{i}{\hbar v_{f}}\tfrac{\chi(x)}{\chi'(x)}$ and $\overline{V}_{23}(x)=\overline{V}_{32}^{*}(x)=i m$. We are thus left with the components
\begin{equation}
\overline{V}_{13}(x)=\frac{\epsilon m}{\hbar v_{f}}\frac{\chi(x)}{\chi'(x)} , \quad 
\overline{V}_{31}(x)=\frac{m}{\hbar v_{f}\epsilon}\frac{-m^{2}\chi(x)+\hbar^{2}v_{f}^{2}\chi''(x)}{\chi'(x)} ,
\end{equation}
which are both real-valued quantities. The requirement of hermiticity $\overline{V}_{13}(x)=\overline{V}_{31}(x)$ is equivalent to a simple differential equation for $\chi(x)$. Keeping in mind that $\chi(x)$ has to be monotonous, we fix $\chi(x)=\sinh(\xi x)$, with $\hbar v_{f}\xi=\sqrt{m^{2}+\epsilon^{2}}$. 

The resulting Hamiltonian takes the form
\begin{equation}
\widetilde{H}=-i\hbar v_{f}S_{1}\partial_{x}+\frac{\tanh(\xi x)}{\hbar v_{f}\xi}\left( m^{2}\widetilde{S}_{1}+\epsilon m S_{2} \right) , \quad \xi=\frac{\sqrt{m^{2}+\epsilon^{2}}}{\hbar v_{f}} ,
\end{equation}
from which only the missing state solution associated with $\epsilon=0$ has a finite norm, which forms the unique bound state
\begin{equation}
\widetilde{\boldsymbol{\Psi}}_{\epsilon=0}\propto 
\begin{pmatrix}
0 \\
i m \\
\epsilon
\end{pmatrix}
\operatorname{sech}(\xi x) .
\end{equation}

\subsection{Case $\Lambda=diag(\epsilon,0,0)$\label{caseIV}}
As a last case of study, let us set the factorization energies
\begin{equation}
\epsilon_1=\epsilon,\quad\epsilon_2=\epsilon_3=0,
\end{equation}
where both seed solutions corresponding to the zero energy are degenerate Bloch states from the flat band. In such a case, we construct the seed matrix $U$ as
\begin{equation}
U(x)=\left( \boldsymbol{\Psi}_{\epsilon}, \boldsymbol{\Psi}_{fb;1}, \boldsymbol{\Psi}_{fb;2}\right)=
\begin{pmatrix}
\sinh(\nu x) & m \chi_{1}(x) & m \chi_{2}(x)  \\
i\frac{\epsilon}{\hbar v_{f}\nu}\cosh(\nu x) & 0 & 0 \\
-\frac{m}{\hbar v_{f}\nu}\cosh(\nu x) & -\hbar v_{f} \chi_{1}'(x) & -\hbar v_{f} \chi_{2}'(x)
\end{pmatrix},
\end{equation}
whose determinant becomes
\begin{equation}
\Delta(x):=\det(U(x))=-i\frac{m\epsilon}{\nu}\cosh(\nu x)\operatorname{Wr}(\chi_{2},\chi_{1}) , 
\label{case-3-det}
\end{equation}
with $\operatorname{Wr}(f,g)=fg'-f'g$ the \textit{Wronskian} of $f$ and $g$. 

Therefore, we shall fix $\chi_{1,2}$ such that $\operatorname{Wr}(\chi_{2},\chi_{1})\neq 0$ for $x\in\mathbb{R}$ to keep $U$ invertible. On the one hand, the new matrix potential $\overline{V}$ has the null components $\overline{V}_{13}(x)=\overline{V}_{33}(x)=0$. On the other hand, the components $\overline{V}_{12}(x)=\overline{V}_{21}^{*}(x)=-i \hbar v_{f} \nu^{2}\tanh(\nu x)$ automatically fulfill the hermiticity condition and are independent of the choice of $\chi_{1,2}(x)$. Likewise, the condition $\overline{V}_{23}(x)=\overline{V}_{32}^{*}(x)$ imposes another condition for $\chi_{1,2}$. From the latter considerations, we obtain the first relation
\begin{equation}
\frac{\operatorname{Wr}(\chi_{2}',\chi_{1}')}{\operatorname{Wr}(\chi_{2},\chi_{1})}=-\frac{m^{2}}{\hbar^{2} v_{f}^{2}},
\label{case-3-cond-1}
\end{equation}
and, given that $\overline{V}_{13}(x)=0$, we must fix $\overline{V}_{31}(x)=0$, leading to the second relation
\begin{equation}
\frac{d}{dx}\ln \frac{\operatorname{Wr}(\chi_{2},\chi_{1})}{\cosh(\nu x)}=\frac{\hbar^{2} v_{f}^{2}\nu}{m^{2}}\tanh(\nu x)\frac{\operatorname{Wr}(\chi_{2}',\chi_{1}')}{\operatorname{Wr}(\chi_{2},\chi_{1})}.
\label{case-3-cond-2}
\end{equation}

The relations in Eqs.~\eqref{case-3-cond-1}-\eqref{case-3-cond-2} provide a set of relationships to determine the unknown functions through
\begin{equation}
\operatorname{Wr}(\chi_{2},\chi_{1})\equiv\chi_{2}\chi'_{1}-\chi_{2}'\chi_{1}=\ell , \quad \operatorname{Wr}(\chi_{1}',\chi_{2}')\equiv\chi_{1}'\chi_{2}''-\chi_{1}''\chi_{2}'=-\ell\frac{m^{2}}{\hbar^{2}v_{f}^{2}} ,
\label{case-3-WR3}
\end{equation}
with $\ell\in\mathbb{R}$ a non-null integration constant so that the determinant~\eqref{case-3-det} is non-null for $x\in\mathbb{R}$; that is, $U$ becomes invertible. 

In order to solve~\eqref{case-3-WR3}, we make use of the ansatz $\chi_{1}(x)=f(x)$ and $\chi_{2}(x)=\eta(x)f(x)$, with $f(x)$ and $\eta(x)$ to be determined. Such an ansatz allows decoupling the set of equations in~\eqref{case-3-WR3} so that we are left with the simpler problem
\begin{equation}
f''-\nu_{0}^{2} f=0 , \quad -\eta'f^{2}=\ell, \quad \nu_{0}^{2}=\frac{m^{2}}{\hbar^{2}v_{f}^{2}} .
\label{case-3-chi}
\end{equation}
Notice that $\chi_{2}(x)=f(x)\eta(x)=-\ell f(x)\int^{x} dx' f^{-2}(x')$ also fulfills the first differential equation in~\eqref{case-3-chi}, i.e., $\chi_{2}(x)$ is the second linearly independent solution whenever $\ell\neq 0$. 

In this form. we only have to find two linearly independent solutions of $f''-\nu_{0}^{2} f=0$ and label them as $\chi_{1}(x)$ and $\chi_{2}(x)$. Among all the possible linearly independent solutions, we particularly use
\begin{equation}
\chi_{1}(x)=\sinh(\nu_{0}x) , \quad \chi_{2}(x)=\frac{\ell}{\nu_{0}}\cosh(\nu_{0}x),
\label{case-3-seed-1}
\end{equation}
where the factor $\nu_{0}^{-1}$ has been included so that we recover the required Wronskian relation $\operatorname{Wr}(\chi_{2},\chi_{1})=\ell$. Any other linear combination of the previous solutions can be used as seed solutions; however, we select (\ref{case-3-seed-1}) for the sake of simplicity. The new Hamiltonian takes the form
\begin{equation}
\widetilde{H}=-i\hbar v_{f} S_{1}\partial_{x} + \hbar v_{f}\nu \tanh(\nu x)\widetilde{S}_{1}-\epsilon\widetilde{S}_{3}\equiv 
\begin{pmatrix}
\mathfrak{H} & \vec{0}^{T} \\
\vec{0} & 0 
\end{pmatrix}.
\label{Hreduced}
\end{equation}
Here, we have introduced the $2\times 2$ Hamiltonian $\mathfrak{H}=-i\hbar v_{f}\sigma_{1}\partial_{x}+\hbar v_{f}\nu\tanh(\nu x)\sigma_{2}-\epsilon\sigma_{3}$, and $\vec{0}=(0,0)$. This is an interesting result, as the interaction of the atoms on the $C$-sites with their neighbors on the $A$- and $B$-sites is effectively zero, partially isolating the lattice interactions. Furthermore, the term $\mathfrak{H}$ renders an effective $2\times 2$ Hamiltonian, which may be associated with the dynamics of a graphene lattice with vector potential $A(x)=\hbar v_{f}\nu \tanh(\nu x)$ under the Landau gauge, a mass term $-\epsilon\sigma_{3}$, and null transverse momentum $k_{y}$. Effectively, we have thus found an intertwining between a reflectionless graphene model and a free-particle Lieb lattice. 

The only square-integrable missing state solution is given by
\begin{equation}
\widetilde{\boldsymbol{\Psi}}_{\epsilon}(x)=
\sqrt{\frac{\nu}{2}}\begin{pmatrix}
0 \\ \operatorname{sech}(\nu x)\\0
\end{pmatrix} ,\quad \widetilde{H}\widetilde{\boldsymbol{\Psi}}_{\epsilon}=\epsilon\widetilde{\boldsymbol{\Psi}}_{\epsilon}
\end{equation}

In particular, a flat-band solution maps as $L\boldsymbol{\Psi}_{fb}=-\hbar v_{f}\left( 0, 0, \chi''-\nu_{0}^{2}\chi \right)^{T}$. That is, flat-band solutions of the Lieb lattice are mapped into trivial solutions of the effective  graphene model $\mathfrak{H}$. This is unsurprising, for the graphene model does not possess a flat-band spectrum. In turn, the map of the Lieb lattice free-particle solutions with eigenvalue $\epsilon'\neq \epsilon$ leads to the new solutions
\begin{equation}
\widetilde{\boldsymbol{\Psi}}_{\epsilon'}=L\boldsymbol{\Psi}_{\epsilon'}=\begin{pmatrix}
\widetilde{\psi} \\ 0 
\end{pmatrix}
, \quad
\widetilde{\psi}_{\epsilon'}=\cosh(\nu'x)
\begin{pmatrix}
\frac{\nu'^{2}-\nu^{2}}{\nu'} \\ i\frac{\epsilon'}{\hbar v_{f}}\left( \tanh(\nu'x)-\frac{\nu}{\nu'}\tanh(\nu x) \right)
\end{pmatrix}.
\end{equation}
Likewise, the $2\times 2$ solution $\phi_{\epsilon}$ is also an eigensolution of the lower-dimensional model for $E=\epsilon'$.

\subsection{Absence of back scattering}
In all the considered examples, the new potentials $\widetilde{V}$ converged to constant matrices for large $|x|$. Therefore, we can expect that the scattering states will asymptotically acquire the form plane waves. The new Hamiltonians were intertwined with the free-particle operator by the intertwining operator $L$ defined via the matrix $U$, see (\ref{L}). Therefore, we can exploit this mapping to find scattering amplitudes. 

The intertwining operator is regular in all the constructed models, i.e., it maps a bounded regular function into another bounded regular function. Let us consider its action on the scattering states of the free particle Hamiltonian (\ref{Dirac-H2}). We take waves coming from the left $\boldsymbol{\Psi}=e^{ik x}(a,b,c)^T$, for some $k>0$. The explicit form of the coefficients  $a,\ b,\ c$ is not important in our analysis, and we leave them defined implicitly. Let us see how is the asymptotic behavior of the transformed function. For large $|x|$, we get
\begin{equation}
L\boldsymbol{\Psi}|_{x\rightarrow\pm\infty}\sim (ik+W_{\pm})\boldsymbol{\Psi},\quad W_{\pm}:=\lim_{x\rightarrow\pm\infty}U'U^{-1}.
\end{equation}
with $W_{\pm}$ a constant matrix that acquires different forms for each of the presented models. Apparently, the mapped state $L\boldsymbol{\Psi}$ keeps the same momentum as $\boldsymbol{\Psi}$, for it does not contain any reflected component. Therefore, the potentials $\widetilde{V}$ of the new systems are {\textit{reflectionless}}.

\section{Concluding remarks}
\label{sec:conclusions}

\begin{table}
\centering
\begin{tabular}{c|c|c|c}
Case & $\Lambda=diag(\epsilon_{1},\epsilon_{2},\epsilon_{3})$ & $U=(\boldsymbol{\Psi}_{\epsilon_{1}},\boldsymbol{\Psi}_{\epsilon_{2}},\boldsymbol{\Psi}_{\epsilon_{3}})$ & Discrete spectrum $\sigma(\widetilde{H})$ \\
\hline
I & $diag(\epsilon,0,-\epsilon)$ & $(\boldsymbol{\Psi}_{\epsilon},\boldsymbol{\Psi}_{fb},S\boldsymbol{\Psi}_{\epsilon})$ & $\{\epsilon,0,-\epsilon\}$ \\
II & $diag(m,0,\epsilon)$ & $(\boldsymbol{\Psi}_{m},\boldsymbol{\Psi}_{fb},S\boldsymbol{\Psi}_{\epsilon})$ & $\{0,\epsilon\}$ \\ 
III & $diag(\epsilon,\epsilon,0)$ & $(\boldsymbol{\Psi}_{\epsilon},\boldsymbol{\Psi}_{fb;1},\boldsymbol{\Psi}_{fb;2})$ & $\{0\}$ \\ 
IV & $diag(0,0,\epsilon)$ & $(\boldsymbol{\Psi}_{0},\boldsymbol{\Psi}_{0},S\boldsymbol{\Psi}_{\epsilon})$ & $\{\epsilon\}$
\end{tabular}
\caption{Summary of the allowed factorization energies, the related seed functions, and the bound states inherited to the new Hamiltonian. Here, $\boldsymbol{\Psi}_{fb;j}$ stands for a flat-band solution with an arbitrary function $\chi_{j}(x)$, together with $\boldsymbol{\Psi}_{0}\equiv \boldsymbol{\Psi}_{\epsilon=0}$, and $\boldsymbol{\Psi}_{m}\equiv \boldsymbol{\Psi}_{\epsilon=m}$.}
\label{tab:summary}
\end{table}

In the manuscript, we have demonstrated that the implementation of Darboux transformation on pseudospin-one systems can substantially benefit from the use of flat-band solutions. We considered  a chiral-symmetric free-particle Lieb lattice Hamiltonian (\ref{Dirac-H2}) that possesses a flat-band solution~\cite{Jak23a,Jak23b} as the initial system. We presented four different examples that differ by the explicit form of the matrix $U$ that defines both the intertwining operator $L$ and the new Hamiltonian $\widetilde{H}$. Each of the cases corresponded to a different conceptual choice of the factorization energies; see Tab.~\ref{tab:summary}. It was also shown that the new models could represent effective interactions caused by inhomogeneous hopping amplitudes in the Lieb lattice, see  Sec~\ref{sec:Lieb}. 

In the Sec.\ref{CaseI} (Case-I), the choice of the factorization energies $\{\epsilon,0,-\epsilon\}$ provided us with the systems that possessed chiral symmetry. We demonstrated in this example that the arbitrariness of the flat-band solution is vital in order to guarantee the Hermiticity of the new Hamiltonian $\widetilde{H}$. This is a fact that, to the best of the authors' knowledge, has not been exploited in the literature.
The model presented in Sec. \ref{caseII} (Case-II) was based on the factorization energies ${m,0,\epsilon}$. This led to a new model with only two bound states, as the missing state $\widetilde{\Psi}_m$ proved to be non-square-integrable. However, a new bound state is generated in this case at the zero-energy level. It is worth noticing that the first two examples are inspired by the choice of factorization energies performed in \cite{Samsonov} in the analysis of Darboux transformation for spin-$1/2$ Dirac systems.

In Sec. \ref{caseIII} (Case-III), two of the three factorization energies were set to an identical, nonvanishing value. Such a choice would lead to a trivial result in the case of pseudospin-$1/2$ systems, as the Darboux transformed Hamiltonian would be equivalent to the original one. We illustrated on the explicit model that this is not the case in spin-one settings. The last example of the Sec. \ref{caseIV} (Case-IV) is particularly remarkable: fixing the two factorization energies to zero and utilizing the flat-band modes as the seed solutions, it was possible to get the new system whose Hamiltonian reduces effectively to the energy operator of spin-$1/2$, see (\ref{Hreduced}). This example opens the way to associate two seemingly different lattice models with different values of pseudospin. The question emerges whether the Darboux transformation could be used to intertwine exactly solvable models of Dirac fermions in graphene with the new solvable models of pseudospin-1 quasi-particles in the Lieb lattice. This is a topic in progress, and results in this regard will be reported elsewhere.


\section*{Acknowledgments}
K.Z. acknowledges the support from the project ``Physicists on the move II'' (KINE\'O II) funded by the \textit{Ministry of Education, Youth, and Sports} of the Czech Republic, Grant No. CZ.02.2.69/0.0/0.0/18 053/0017163.

\appendix
\setcounter{section}{0}
\renewcommand{\thesection}{A-\arabic{section}}
\renewcommand{\theequation}{A-\arabic{equation}}
\setcounter{equation}{0}  


\end{document}